\documentclass[12pt]{article}
\usepackage{amssymb}
\usepackage{amsmath}
\usepackage{graphicx}

\begin{document}

\vspace*{1cm}

\begin{center} 
\setlength{\baselineskip}{24pt}
{\LARGE Heisenberg and the Levels of Reality
}
\end{center}

\begin{center}
\vspace{2cm}
{\large Basarab Nicolescu}

\vspace{0.5cm} 
 Theory Group,  Laboratoire de Physique Nucl\'eaire  et des Hautes  \'Energies 
(LPNHE)\footnote{Unit\'e  de Recherche des Universit\'es 
  Paris 6 et Paris 7, Associ\'ee ou CNRS},
 CNRS and Universit\'e Pierre et Marie Curie, Paris\\ 
     {\small e-mail: \texttt{nicolesc@lpnhep.in2p3.fr }}

\vspace{3cm}
\textbf{Abstract}
\end{center}
We first analyze the transdisciplinary model of Reality and its key-concept of "Levels of Reality". We then compare this model with the one elaborated by Werner Heisenberg in 1942.

\newpage

\section{Introduction}
The idea of 'Levels of Reality' came to me during a post-doctoral visit at Lawrence Berkeley Laboratory, in 1976. I did not understand from where was coming the resistance to the unification between the relativity theory and the quantum mechanics. This was the starting point of my reflection. At that time, I was working with Geoffrey Chew, the founder of bootstrap theory. The discussions we had together and with other colleagues from Berkeley, have stimulated me to formulate this idea. It is at Berkeley that I have begun writing a book regarding the epistemological and philosophical extensions of the quantum physics. 

In 1981, I was intrigued by the notion of 'veiled Real' by Bernard dÕEspagnat 
\cite{espa81}, that did not seemed to me to be a satisfying solution to the
problems I was dealing with and I decided to make public my notion of
'Levels of Reality'. Therefore, I introduced this notion into an article
published in 1982 \cite{nico82}. The form of this concept was resumed in the
first edition of my book \textit{Us, the Particle and the World}
\cite{nico85}. Afterwards, during the years, I developed this idea in several books, articles and conferences.

In 1992, I was invited as an expert to the plenary session of the Pontifical Academy of Sciences dedicated to the study of complexity in sciences. I spoke on Nature considered from the quantum physics point of view and I presented my approach concerning the Levels of Reality \cite{nico96}. The Austrian physicist Walter Thirring, present at the congress at Vatican, gave me a little article, unpublished yet, where I have discovered his important considerations on the nature of physical laws, in the case of different Levels of Reality \cite{thir95} .

But the big surprise came in 1998, when I discovered the work of Werner
Heisenberg \textit{Philosophy : The Manuscript of 1942} \cite{heis42}. This text has provoked in me a veritable astonishment because I found the same idea of Levels of Reality, obviously under a different form. HeisenbergÕs book had an amazing history: it was written in 1942, but it was published in German only in 1984. It was translated in French in 1998. As far as I know, there is no English translation of this work. 

The opinion that I want to express in this paper is in total agreement with those of the quantum mechanics founders: Werner Heisenberg, Wolfgang Pauli and Niels Bohr, but due to space reasons, I will treat in the present study only the philosophical ideas of Heisenberg. I shall start by exposing my own ideas, to continue by studying the correspondence that exists between them and those of Heisenberg.

\section{Classical realism and quantum realism}

The modern science is founded on the idea of a total separation between the observing-subject and the Reality, assumed to be completely \textit{independent} from the first one. But, at the same time, in the modern science are given three fundamental postulates, which are extending at a supreme degree the research of laws and order:
\begin{itemize}
  \item ¥	The existence of universal laws, with mathematic character.
  \item ¥	The discovery of these laws by the scientific experiment.
  \item ¥	The perfect reproducibility of the experimental data.
\end{itemize}
The extraordinary success of the classical physics, from Galileo, Kepler and Newton until Einstein, have confirmed the validity of these three postulates. At the same time, they have contributed to the instauration of a \textit{simplicity} paradigm, which became dominant during the XIXth century.

The classical physics is founded on the idea of \textit{continuity}, in agreement with the evidence supplied by the sense organs: we canÕt pass from one point of the space and of the time to another, without passing through all intermediary points.

The idea of continuity is intimately linked to a key concept of the classical physics: the \textit{local causality}. Every physical phenomenon could be understood by a continuous chain of causes and effects: to every cause at a certain point corresponds an effect to an infinitely near point and to every effect at a certain point corresponds a cause to an infinitely near point. There is no need of any direct action at distance.

The concept of \textit{determinism} is central in the classical physics. The classical physics equations are such that if one knows the positions and the velocities of the physical objects at a certain moment, one can predict their positions and velocities at any other moment of time. The laws of classical physics are deterministic laws. The physical states being functions of position and velocity, it results that if the initial conditions are known (the physical state at a given moment of time) one can \textit{completely} predict the physical state at any other moment of time.  

\textit{The objectivity} of the classical physics is fundamentally linked to the knowledge of an object moving in the 1-dimensional time and the 3-dimensional space. The central role of the space-time in four dimensions was not altered by the two relativity theories of Einstein, restricted and general, that constitute the apogee of the classical physics.

The quantum mechanics is in a total conceptual rupture with the classical mechanics.

According to Planck's discovery, the energy has a discontinuous, discrete structure. The \textit{discontinuity} means that between two points there is nothing, no objects, no atoms, no molecules, no particles, just \textit{nothing}. And even the word 'nothing' is too much.

A physical quantity has, in quantum mechanics, several possible values associated with given probabilities of occurrence. But in a physical measurement we get obviously just one single result. This abolition, via the measurement process, of the plurality of possible values of an observable quantity had an obscure meaning but it already clearly indicated the existence of a new type of causality.

Seven decades after the quantum mechanics was born, the nature of this new
type of causality was clarified thanks to a rigorous theoretical result, the
Bell's theorem, and also to high precision experiments. A new concept made
in this way its entrance in physics: \textit{the non-separability}. The
quantum entities continue to interact, never mind the distance between them.
Therefore, a new type of causality appears -  \textit{global causality} - that concerns the system of all physical entities, in their ensemble.  

The quantum entities, the 'quantons', are at the same time corpuscles and waves or, more precisely, they are neither corpuscles nor waves. 

The famous uncertainty relations of Heisenberg show without any ambiguity that is impossible to localise a quanton into an exact point of the space and an exact point of time. In other words, it is impossible to assign a well-determined trajectory to a quantum particle. The \textit{indeterminism}, reigning at the quantum scale, is a structural indeterminism, fundamental and irreducible. It does not means neither hazard nor imprecision.

The so-called \textit{quantum paradoxes} (as, for example, the famous
paradox of 'Schroedinger's cat') are false paradoxes, because they point out
contradictions exclusively in correlation with the natural, ordinary language, which is that of the classical realism: these end to be paradoxes when the language appropriate to the quantum mechanics is used. Even if they are instructive when one wants to show the incompatibility between the classical and quantum realism, these paradoxes become useless in the context of the quantum ideas. 

The true question is the incompatibility between the classical realism and the quantum one. 

The classical object is localised in space-time while the quantum object is not localised in space-time. It moves into an abstract mathematical space, ruled by the algebra of operators and not by the algebra of numbers. In quantum physics, the abstraction is no longer a simple tool to describe reality but a constitutive part of reality itself. 

The classical object is subjected to the local causality, while the
quantum object is not submitted to this causality. It is impossible to
predict an individual quantum event. One can predict only the occurrence
probabilities of the events. The key of understanding this seemingly
paradoxical and also irrational (from the point of view of classical
realism) situation is the quantum superposition principle: the superposition of two quantum states is also a quantum state. 

It is impossible to obtain the classical mechanics as a particular case of
the quantum mechanics because the h constant, characterising the quantum
interactions, the famous Planck constant, has a well-determined value. This
value is different from zero. The limit h going to 0 has no rigorous meaning. 

The radical break between the classical and quantum realism explain why one had not succeed until now to unify the theories of relativity and of the quantum mechanics into a single one, despite the fulminating evolution of the quantum field theory resulting in the superstrings theory.

It is even possible that such a unifying theory will never be found. Does this incompatibility mean that we have reached a limit in the physical description of reality, or that a new characteristic of reality is to be discovered?  It is this second possibility that I want to explore now.

\section{The Levels of Reality}

I interpreted the incompatibility between the quantum mechanics and the classical mechanics as meaning the necessity of enlarging the domain of reality, by abandoning the classical idea of existence of only one level of reality.

Let us give to the word 'reality' its pragmatic and ontological meaning.

I understand by Reality, everything that resists to our experiences, representations, descriptions, images or mathematical formalisms. In quantum physics, the mathematical formalism is inseparable from experiment. It resists, in its manner, both by the care for the internal selfconsistence and by the need to integrate the experimental data without destroying this selfconsistence.

We also have to give an ontological dimension to the notion of Reality.

The Nature is an immense and inexhaustible source of the unknown, that justifies the very existence of science. Reality is not only a social construction, the consensus of a community, an intersubjective agreement. It has also a \textit{trans-subjective} dimension, to the extent where a simple experimental fact could ruin the most beautiful scientific theory.

 Of course, I make the distinction between \textit{Real} and \textit{Reality}. \textit{Real} means \textit{what it is}, while \textit{Reality} is connected to the \textit{resistance} in our human experience. The real is, by definition, veiled forever, while the Reality is accessible to our knowledge.

I define a \textit{Level of Reality} as an ensemble of systems invariant to the action of a number of general laws: for example, the quantum entities submitted to the quantum laws, which are on radical break with the laws of the macrophysical world. This means that two levels of Reality are \textit{different} if, passing from one to another, there is a break of the laws and break of the fundamental concepts (as causality, for example).  

The \textit{discontinuity} present in the quantum world is also present in the structure of the Levels of Reality, by the coexistence of macrophysical world and the microphysical world.

The Levels of Reality are radically different from the organisation levels, as they were defined in the systemic approaches. The organisation levels do not suppose a rupture of fundamental concepts: a certain number of organisation levels belong to only one and the same Level of Reality. There is no discontinuity between the organisation levels belonging to a well-determined Reality level. The organisation levels correspond to different arrangements of the same fundamental laws, while the Levels of Reality are generated by the coherent action of radically different ensembles of laws.

The Levels of Reality and the organisation levels offer the possibility of a new taxonomy for the eight thousand academic disciplines existing now. Many disciplines could coexist at an only and the same Level of Reality even if they correspond to different organisation levels. For example, the Marxist economy and the classical physics belong to the only and same level of reality, while quantum physics and psychoanalysis belong to another Level of Reality.  

Due to the notion of Levels of Reality, the Reality acquires a multidimensional and multireferential structure. The Levels of Reality also allow defining useful notions as: levels of language, levels of representation, levels of materiality or levels of complexity.

The Reality comports, according to my approach, a certain number of levels. In fact, the previous considerations concerning two Levels of Reality could be easily generalised to a larger number of levels. The following analysis does not depend on the fact that this number is finite of infinite. For the terminological clarity's sake, I shall assume that this number is infinite.  

Obviously, there is coherence between the different levels of Reality, at
least in the natural world. In fact, a vast \textit{selfconsistence} seems
to rule the evolution of the Universe, from the infinitely small to the infinitely large, from the infinitely short to the infinitely long. For example, a very small variation of the coupling constant of the strong interactions between quantum particles would lead, at the infinitely large scale (our Universe), either to the conversion of all hydrogen in helium, or to the inexistence of complex atoms as the carbon. Or a very small variation of the gravitational coupling constant would lead either to ephemeral planets, or to the impossibility of their formation. Furthermore, according to the actual cosmological theories, the Universe seems able to \textit{create itself} without any external intervention. An information flux is transmitted in a coherent manner from a Level of Reality to another level of Reality of our physical Universe.

Every Level of Reality has its own associated space-time. Thus, the classical Level of Reality is associated to the four dimensional space-time, while the quantum Level of Reality is associated with more than four dimensions. In the most sophisticated and the most promising theory for the unification of all physical interactions Ð the M theory ('M' from 'membrane'), the space-time must have eleven dimensions: one time-dimension and ten space-dimensions. 

The superstrings modify in an interesting manner our conception on the physical reality. The superstring, fundamental entity of the new theory, is an object \textit{spread} in space. Consequently, it is logical impossible to define \textit{where} and \textit{when} are interacting the superstrings. This characteristic is in the spirit of quantum mechanics. On the other hand, their finite dimension implies that there is a \textit{limit} of our possibility to explore reality. Our anthropomorphic convention of distance is no longer applicable. Neither the Universe nor any of its objects have any meaning over this limit. Finally, the space dimensions are of two kinds: large, vast, visible (as the three dimensions of what we consider as our own space) and small, \textit{wrapped} on themselves, invisible.  

A new \textit{Relativity Principle} emerges from our model of Reality:
\textit{no Level of Reality constitutes a privileged place from where one
could understand all the other Levels of Reality}. A Level of Reality is
what it is because all the other levels simultaneously exist. In other words, our model is non-hierarchical. There is no fundamental level, but the absence of fundaments does not mean an anarchical dynamics. The fundaments are replaced by the unified and coherent dynamics of all Levels of Reality, which are already discovered or will be discovered in the future. 

Every Level of Reality is characterised by its \textit{incompleteness}:
the laws ruling this level are just a part of the ensemble of laws ruling
all the Levels of Reality. This property is in agreement with the Goedel
theorem, concerning the arithmetic and all mathematical theory containing
the arithmetic. The Goedel theorem tells that a rich enough axioms system has either undecidable or contradictory results.

The dynamics of the Levels of Reality is made clear in a pertinent manner by three thesis formulated by the physicist Walter Thirring \cite{thir95}:
\begin{itemize}
  \item ¥	\textit{The laws of any inferior level are not completely determined by the superior level laws}. Thus, well-anchored notions in the classical thinking, as 'fundamental' and 'accidental', must to be re-examined. What is considered fundamental at a certain level may appear as accidental at a superior level and what is considered accidental or incomprehensible at a certain level could appear as fundamental at a superior level.
  \item ¥	\textit{The laws of an inferior level depend more of their emergency circumstances than of the superior level laws}. The laws of a certain level essentially depend of the local configuration that they are referred at. Therefore, there is a kind of local autonomy for the respective level of Reality. But certain internal ambiguities concerning the inferior level laws are solved by the consideration of superior order laws. The self consistence of these laws reduces the laws ambiguity.
  \item ¥	\textit{The laws hierarchy advanced in the same time with the Universe itself.} In other words, we assist at the birth of laws as the Universe develops. These laws pre-existed at the 'beginning' of the Universe as possibilities. It is the evolution of the Universe that actualises these laws and their hierarchy.  
\end{itemize}  

  	The zone between different levels of Reality and the zone beyond all
levels of Reality is in fact \textit{a zone of non-resistance} for our
experiments, representations, descriptions, images or mathematical
formalisations. This transparency zone is due to our body and sense organs
limitations, no matter what measurement instruments are prolonging these
sense organs. Therefore, we have to deduce that the 'distance' (understood
as topological distance) between the extreme levels of Reality is
\textit{finite}. But this finite distance does not mean a finite knowledge.
Exactly as a straight-line segment contains an infinite number of points,
the finite topological distance could correspond to an infinite number of Levels of Reality.

	The \textit{Object} is defined, in our model, by the ensemble of Levels of Reality and its complementary non-resistance zone.

We see therefore, all the difference between my approach of Reality and
that of Bernard d'Espagnat. For d'Espagnat it is in fact only one level of
reality, the empirical reality, surrounded by a diffuse zone of non-resistance, which corresponds to the veiled Real. The veiled Real, by definition, do not resist. Consequently, it does not have the characteristics of a level of Reality.  

Inspired by the phenomenology of Edmund Husserl \cite{huss66}, I assert that the different Levels of Reality are accessible to the human knowledge due to the existence of different \textit{levels of perception}, which are in biunivoque correspondence with the Levels of Reality. These levels of perception allow a more general, unifying, inclusive vision of the Reality, without exhausting it. The coherence of the levels of perception assumes, as in the case of Levels of Reality, a zone of \textit{non-resistance} to the perception.

The ensemble of levels of perception and its complementary zone of non-resistance constitute, in our approach, the \textit{Subject}. 

The two zones of non-resistance, of the Object and of the Subject, must to be \textit{identical} in order to have an information flux able to circulate in a coherent manner between the Object and the Subject. This zone of non-resistance corresponds to a third Interaction term between the Subject and the Object, which could not be reduced neither to the Object nor to the Subject.

Our ternary partition - Subject, Object, Interaction -  is, of course,
different from the binary partition - Subject, Object - of the classical realism.

\section{Heisenberg's model}

Now, I want to analyse the correspondence between my ideas and those of
Werner Heisenberg (1901-1976), expressed in his \textit{Manuscript of 1942}.

As written in the excellent introduction to this book (Ref. 6, p. 17), the axe of the philosophical thinking of Heisenberg is constituted by "two directory principles: the first one is that of the division in Levels of Reality, corresponding to different objectivity modes depending on the incidence of the knowledge process, and the second one is that of the progressive erasure of the role played by the ordinary concepts of space and time." (Ref. 6, p. 240)

For Heisenberg, the reality is "the continuous fluctuation of the experience as gathered by the conscience. In this respect, it is never wholly identifiable to an isolated system" (Ref. 6, p. 166). The reality could not be reduced to substance. For us, the physicists of today, this is evident: the matter is the complexus substance-energy-space-time-information.

As written by Catherine Chevalley, "the semantic field of the word reality included for him everything given to us by the experience taken in its largest meaning, from the experience of the world to that of the soul's modifications or of the autonomous signification of the symbols." (Ref. 6, p. 145)  

Heisenberg does not speak in an explicit manner about 'resistance' in
relation with reality, but its meaning is fully present: "the reality we can
talk about - writes Heisenberg - is never the reality 'in itself', but only
a reality about which we may have knowledge, in many cases a reality to
which we have given form." (Ref. 6, p. 277) The reality being in constant
fluctuation, all we can do is to understand partial aspects thanks to our
thinking, extracting processes, phenomena, and laws. In this context, it is
clear that one can not have completeness: "We never can arrive at an exact
and complete portrait of reality"  (Ref. 6, p. 258) - wrote Heisenberg. The
incompleteness of physics laws is hereby present at Heisenberg, even if he
does not make any reference to Goedel's theorems. For him, the reality is given as 'textures of different kind connections', as 'infinite abundance', without any ultimate fundament. Heisenberg states ceaselessly, in agreement with Husserl, Heidegger and Cassirer (whom he knew personally), that one has to suppress any rigid distinction between Subject and Object. He also states that one has to end with the privileged reference on the outer material world and that the only approaching manner for the sense of reality is to accept its division in regions and levels.

The resemblance with my own definition of Reality is striking.

Heisenberg distinguishes `regions of reality' (\textit{der Bereich der Wirklichkeit}) from `levels of reality' (\textit{die Schicht der Wirklichkeit}). 

"We understand by `regions of reality' - writes Heisenberg - [...] an ensemble of nomological connections. These regions are generated by groups of relations. They overlap, adjust, cross, always respecting the principle of non-contradiction."  (Ref. 6, p.  257)

The regions of reality are, in fact, strictly equivalent to the levels of organization of the systemic thinking.

Heisenberg is conscious that the simple consideration of the existence regions of reality is not satisfactory because they will put on the same plan the classical and the quantum mechanics. It is the essential reason that leads him to regrouping these reality regions in different Levels of Reality. His motivation is therefore identical with mine.
 
Heisenberg regroups the numerous regions of reality in three distinct levels.
 
"It is clear - writes Heisenberg - that the ordering of the regions has to substitute the gross division of world into a subjective reality and an objective one and to stretch itself between these poles of subject and object in such a manner that at its inferior limit are the regions where we can completely objectify. In continuation, one has to join regions where the states of things could not be completely separated from the knowledge process during which we are identifying them. Finally, on the top, have to be the Levels of Reality where the states of things are created only in connexion with the knowledge process." (Ref. 6, p. 372)
 
	Heisenberg's approach is compatible with the Relativity Principle present in my approach. Catherine Chevalley underlines that Heisenberg suppresses the rigid distinction between "exact sciences of the objective real world and the inexact sciences of the subjective world" and he refuses "any hierarchy founded on the privilege of certain nomological connexion forms, or on a region of the real considered more objective than the others" (Ref. 6, p. 152). 
 
	The first Level of Reality, in the Heisenberg model, corresponds to the states of things, which are objectified independently of the knowledge process. He situates at this first level the classical mechanics, the electromagnetism and the two relativity theories of Einstein, in other words the classical physics. 
 
The second Level of Reality corresponds to the states of things inseparable from the knowledge process. He situates here the quantum mechanics, the biology and the consciousness sciences.
 
Finally, the third Level of Reality corresponds to the states of things created in connexion with the knowledge process. He situates on this Level of Reality philosophy, art, politics, 'God' metaphors, religious experience and inspiration experience.  
  
If the first two Levels of Reality of Heisenberg entirely correspond to my own definition, his third level seems to me to mix levels and non-levels (i.e. non-resistance zones). In fact, philosophy, art and politics represent academic disciplines, which are conforming to the intrinsic resistance of a Level of Reality. Even the 'God' metaphors, if they are integrated to a theology, could correspond to a Level of Reality: theology is, after all, a human science as the other ones. But the religious experience and the inspiration experience are difficult to assimilate to a Level of Reality. They rather correspond to the passage between different Levels of Reality in the non-resistance zone. 
  
Obviously, there is an important difference between the two definitions of the Level of Reality notion. The absence of resistance and the absence of the discontinuity in the HeisenbergÕs definition explain this difference. 
  
"The concepts are, so to say, the privileged points where the different Levels of Reality are interweaving" Ð wrote Heisenberg. He specifies on:  "When one is questioning the nomological connexions of reality, these last ones are found every time inserted into a determined reality level; it could not at all be interpreted differently from the concept of reality 'level' (it is possible to speak about the effect of a level onto another one only by using very generally the 'effect' concept). On the other hand, the different levels are connected in the associated ideas and words and which, from the beginning, are in simultaneous relation with the numerous connexions". (Ref. 6, p. 257) This is vague enough and necessarily introduces confusion between the organisation levels and the levels of Reality. If the levels are 'interweaving', one canÕt understand how is possible to introduce a classification of the levels of Reality. The nomological connexions characterise as well the reality regions and the levels of Reality. Therefore, they are not sufficient in order to distinguish 'region' from 'level'.  
  
In fact, Heisenberg does not explicitly impose the non-contradiction principle that could lead him to the discovery of the Levels of Reality discontinuity. However, the discontinuity is mentioned a few times in the \textit{Manuscript of 1942} but only in relation with history: history of representations, history of the individual, history of humanity.
  
Heisenberg also insists on the role of the intuition : "Only the intuitive
  thinking Ð wrote Heisenberg Ð can pass over the abyss that exists between
  the concepts system already known and the new concepts system; the formal
  deduction is helpless on throwing a bridge over this abyss." (Ref. 6, p. 261) But Heisenberg does not draw the logical conclusion that is imposed by the helplessness of the formal thinking: only the non-resistance of our experiences, representations, descriptions, images or mathematical formalisations could bring a bridge over the abyss between two zones of resistance. The non-resistance is the key of understanding the discontinuity between two immediately neighbour levels of Reality.  
   
But this important difference between the two definitions of the Levels of Reality, that of Heisenberg and mine, do not erase the motivation of introducing these levels, motivation which is identical in both cases. 
   
In order to finish, I want to make a few brief considerations on the
   political and intellectual context in which was written \textit{The
   Manuscript of 1942}. During the Nazism, the anti-Semitism included the
   attacks against the relativity theory and the quantum mechanics, viewed
   as products of the Western decadence. The promoters of \textit{Deutsche
   Physik} presupposed that there is only one level of reality. It is
   strange to find as leaders of \textit{Deutsche Physik} two remarkable
   physicists: Philipp Lenard (Nobel prize in 1905) and Johannes Stark
   (Nobel prize in 1919). The anticonceptualism of the \textit{Deutsche
   Physik} holders, at war against 'the Jewish physics', was virulent. The
   knowledge had to discover the unique reality by a language near to the
   intuition. According to them, the experiment was in front of the theory. Their black sheep was the abstract space of the quantum events, while, for them, the only space-time could be that of our ordinary experience, with four dimensions.
   
It is otherwise interesting to find the attachment of totalitarian systems to the four dimensional space-time. There is an astonishing passage in \textit{Materialism and empiriocriticism} (1909), where Lenin attacked the physical theories implying a multidimensional space-time, proclaiming that one can make revolutions only in four dimensions. We may also add here that the notion of Levels of Reality is also mining the fundaments of the dialectical materialism.  
    
Catherine Chevalley was right to write: "\textit{the Manuscript of 1942} appeared as an effort to make \textit{philosophically impossible} an ideological operation as that of \textit{Deutsche Physik}" (Ref. 6, p. 94). The manuscript has circulated among the German physicists and students. To speak about Levels of Reality in the context of the \textit{Deutsche Physik} fight against 'Jewish physics' was equivalent with a true act of resistance against the national-socialism.

 \end{document}